\documentclass[12pt]{iopart}

\usepackage{iopams}  
\usepackage{graphicx}  
\begin{document}

\title[Soft probes of high density QCD physics with CMS]{Soft probes of high density QCD physics with CMS}

\author{Ferenc Sikl\'er for the CMS Collaboration}

\address{KFKI Research Insitute for Particle and Nuclear Physics, Budapest, Hungary}
\ead{sikler@rmki.kfki.hu}
\begin{abstract}
The CMS heavy-ion program will probe QCD matter under extreme conditions. Its
capabilities for the study of global observables and soft probes are
described.
\end{abstract}


\section{Introduction}

The CMS experiment at the LHC is a general purpose detector designed to
explore physics at the TeV energy scale. It has a large acceptance and
hermetic coverage. The various subdetectors are: a silicon tracker with pixels
and strips ($|\eta|<2.4$); electromagnetic ($|\eta|<3$) and hadronic
($|\eta|<5$) calorimeters; and muon chambers ($|\eta|<2.4$).  The acceptance is
further extended with forward detectors: CASTOR ($5.3<|\eta|<6.6$) and Zero
Degree Calorimeters ($|\eta|>8.3$).  CMS detects leptons and both charged and
neutral hadrons. In the following the CMS soft physics capabilities are
described. For a recent extensive review see Ref.~1.

\section{Soft physics}

\subsection{Minimum bias trigger}

To maximize efficiency, it is desirable to trigger on regions with a large
number of produced particles. The level-1 trigger read-out will utilize
hadronic cells, towers. In order to reject non-collision events, such as
those due to beam-gas collisions, the interaction trigger could require a
coincidence of signals from both hadronic forward (HF) calorimeters. The
rejection of such spurious events would not be achievable using only the
central calorimeter as a trigger. At the same time, this choice would reduce
the efficiency for single diffractive processes.

\begin{figure}[t]
 \begin{minipage}[c]{0.45\textwidth}
 \includegraphics[width=\textwidth,angle=-90]{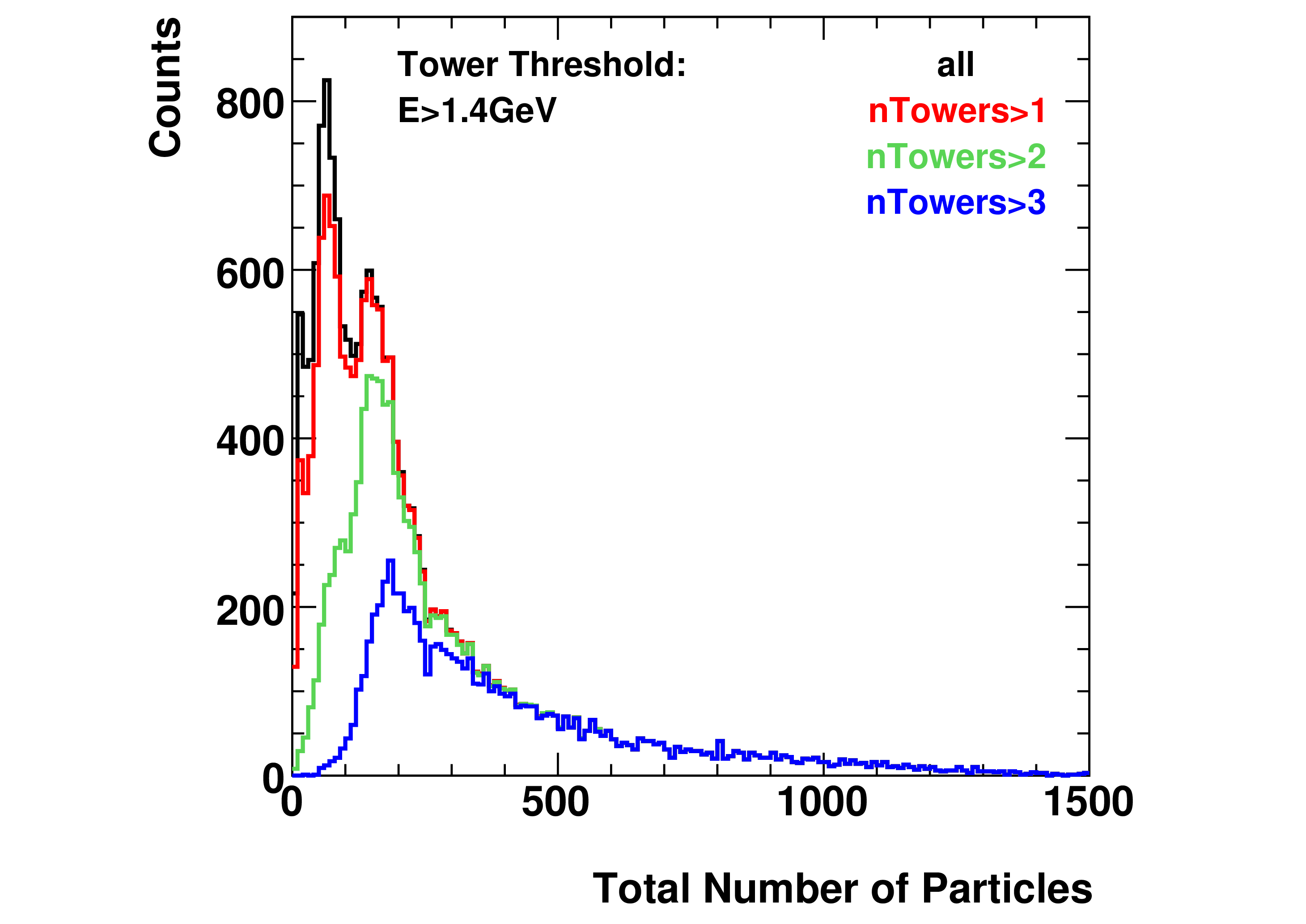}
 \end{minipage}
 \hspace{0.02\textwidth}
 \begin{minipage}[c]{0.45\textwidth}
 \includegraphics[width=\textwidth,angle=-90]{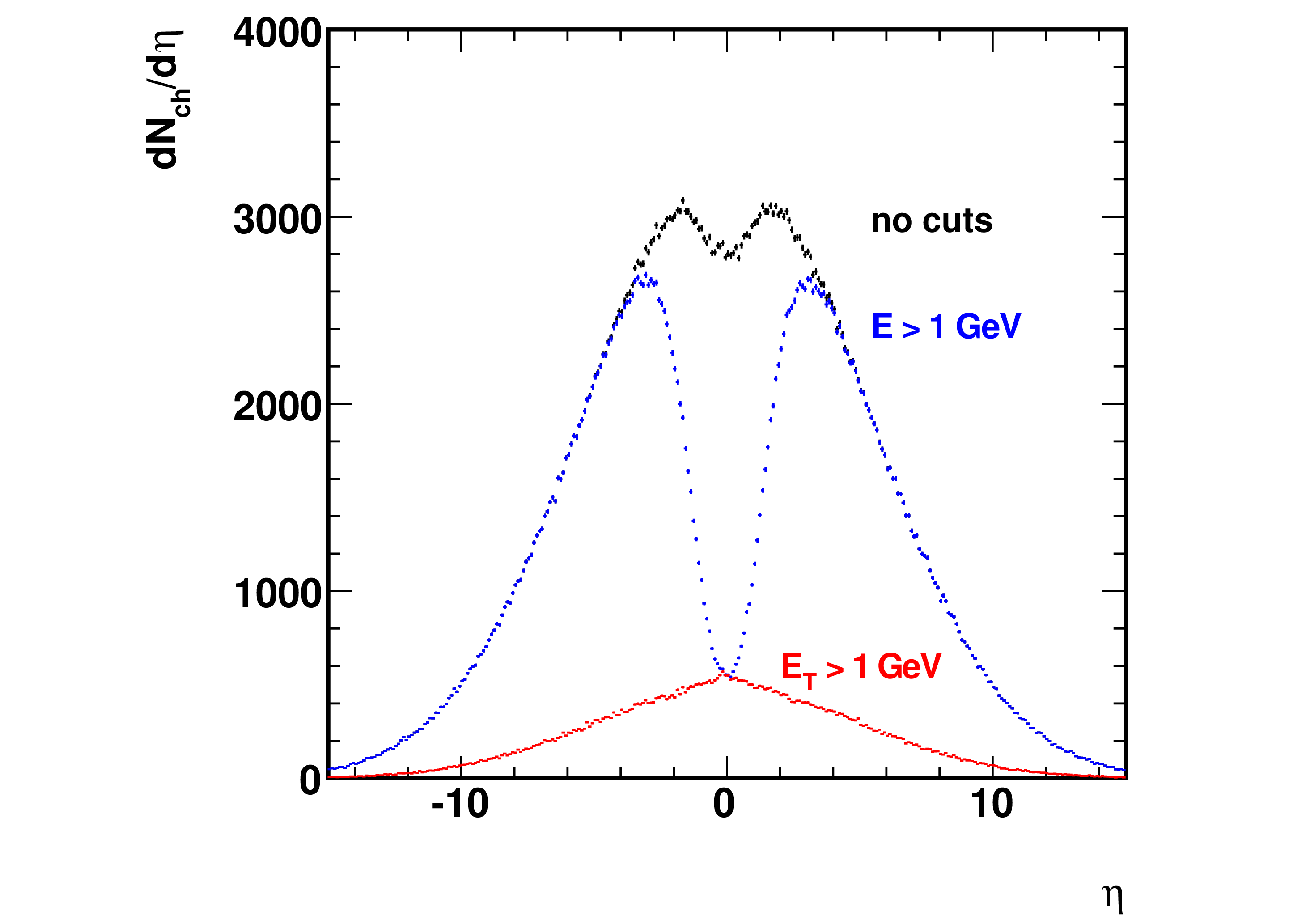}
 \end{minipage}

 \caption{Left: Estimated loss of low multiplicity events due to triggering
requirements on the number of towers for cuts on $E$ in minimum bias p-p
collisions at 14~TeV. Right: Pseudorapidity distribution of charged hadrons in
central Pb-Pb collisions at 5.5~TeV per nucleon from the {\sc hydjet}
\cite{Lokhtin:2005px} generator.  Particle selection to mimic the level-1
trigger is applied for total $\langle E \rangle$ and transverse $\langle E_T
\rangle$ energy.}

 \label{fig:hftrig}
\end{figure}

The p-p trigger will be based on counting towers with energy above the detector
noise level, in both forward hadronic calorimeters ($3<|\eta|<5$). A minimal
number of hits (1, 2 or 3) will be required on one or on both sides.  The
trigger efficiency for different total number of particles using several
symmetric settings is shown in Fig.~\ref{fig:hftrig}-left, where a threshold
value of 1.4~GeV was used. In case of higher luminosity, events with random
trigger could also be taken, because overlapping events can also be
reconstructed.

The Pb-Pb trigger will be similar since there are many particles produced in
the region of the forward calorimeters (Fig.~\ref{fig:hftrig}-right).
Although triggering will be possible for the most central collisions with
transverse energy cuts ($E_T$), more peripheral collisions with lower
multiplicity will suffer from inefficiency. For total energy cuts ($E$), only
events with $b>12$ fm suffer from inefficiency, relative to the p-p baseline.

\subsection{Centrality determination}

\begin{figure}[t]
 \begin{minipage}[c]{0.50\textwidth}
 \includegraphics[width=\textwidth]{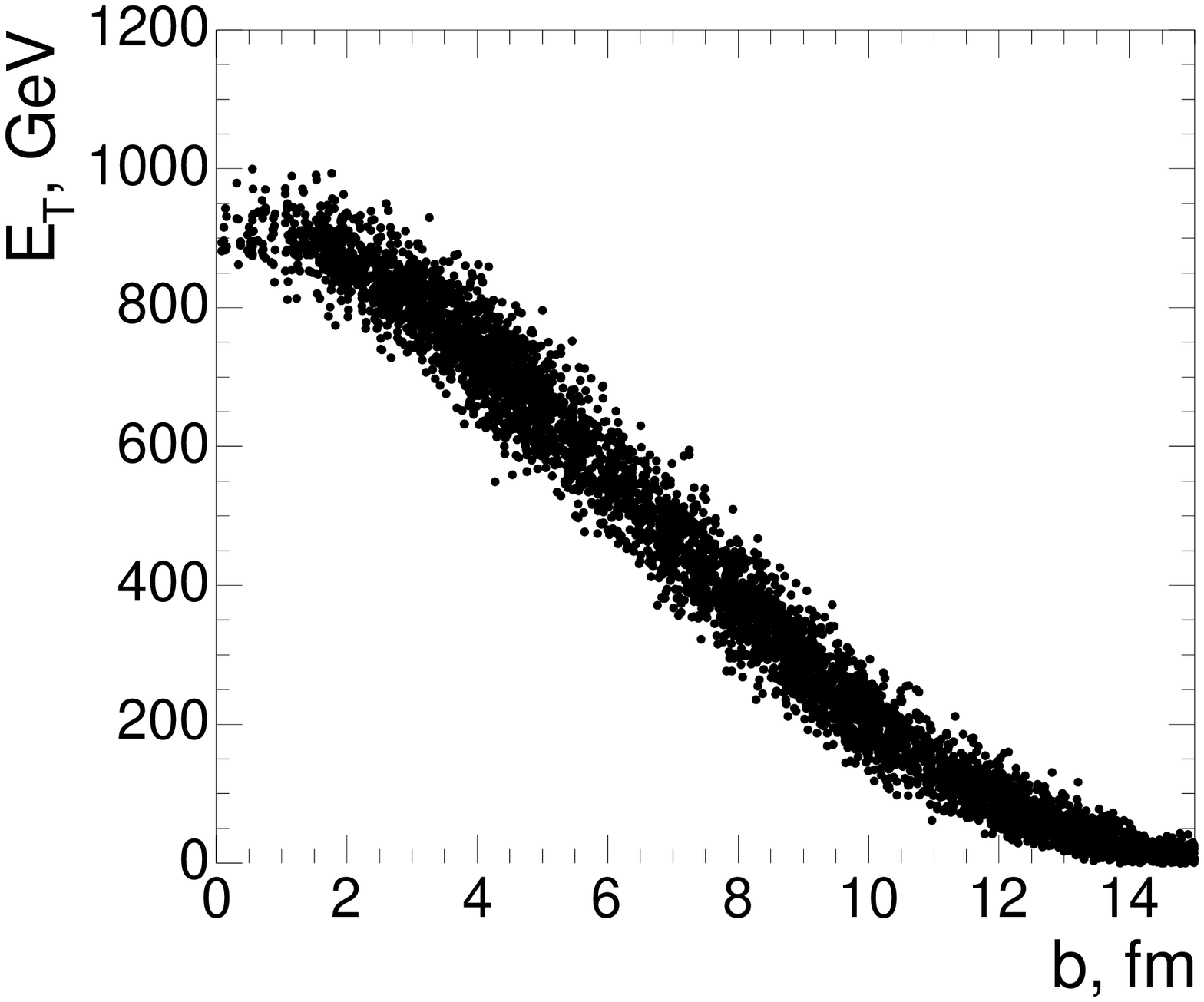}
 \end{minipage}
 \hspace{0.02\textwidth}
 \begin{minipage}[c]{0.45\textwidth}
 \includegraphics[width=\textwidth,angle=-90]{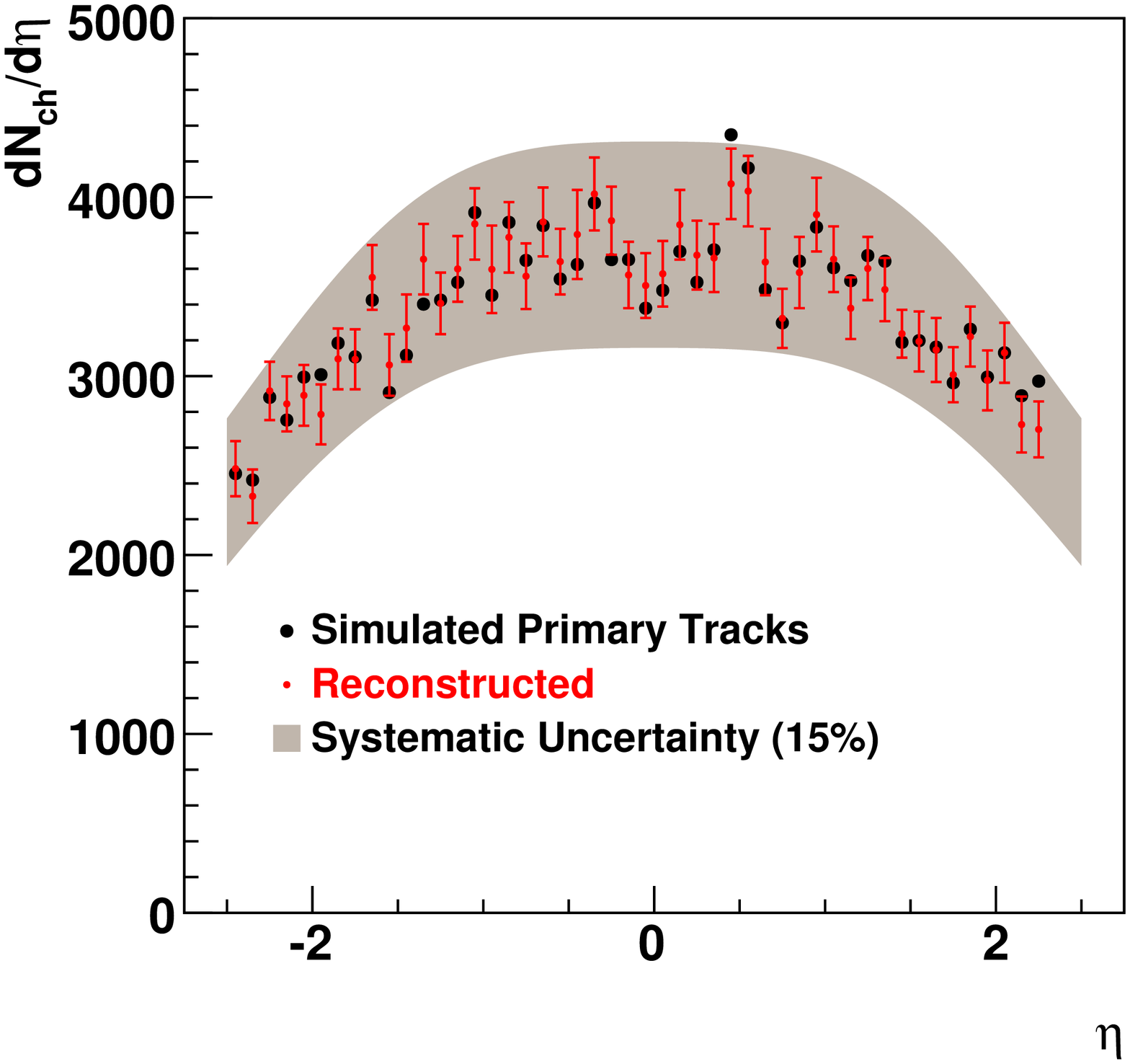}
 \end{minipage}

 \caption{Left: Correlation between the impact parameter, $b$, and the
transverse energy, $E_{\rm T}$, deposited in the forward rapidities, $5.3 <
|\eta| < 6.6$, covered by the  CASTOR calorimeter, as simulated with {\sc
hijing} for 3000 minimum bias Pb-Pb collisions. Right: Comparison of the
original distribution of primary simulated tracks (large points) to the
estimate obtained from the reconstructed hits in the innermost pixel layer
(smaller points with statistical error bars).}

 \label{fig:centmulti}
\end{figure}

In CMS, a simple method of determining the impact parameter on an
event-by-event basis is to use the transverse energy deposited in the
calorimeters, which decreases strongly from central to peripheral collisions.
Due to its relatively low initial parton density, the very forward rapidity
region covered by the HF and CASTOR calorimeters, $|\eta| > 3$, is expected
to be nearly free of final-state rescattering compared to the central
rapidity region. Therefore the energy deposition is determined mainly by the
initial nuclear geometry of the collision rather than by final-state dynamics
(Fig.~\ref{fig:centmulti}-left). Using the forward spectator neutron energy
measured in both zero degree calorimeters can improve the experimental impact
parameter resolution of a few tenths of a fermi.

\subsection{Charged particle multiplicity}

It is possible to measure the rapidity density and multiplicity of produced
charged particles event-by-event. The technique is based on the relation
between the pseudorapidity distribution of reconstructed hit clusters in the
innermost layer of the pixel tracker and that of charged particle tracks
originating from the primary vertex. Corrections are needed for hits from
non-primary sources: looping particles in magnetic field, secondaries and noise. 
Because the amount of energy deposited is proportional to the length
traversed in the silicon, charged particle tracks with longer path lengths
will deposit more energy. This observation is a powerful method for
eliminating most non-primary hits. A cut in $\cosh{\eta}$ selects only hits
where the energy deposition is consistent with particles emanating from the
true collision vertex. Based on previous measurements and recent studies, an
accuracy of about 2\% is expected with systematic errors below 10\%
\cite{hiptdr} (Fig.~\ref{fig:centmulti}-right).

\subsection{Low $p_T$ tracking}

In CMS, the measurement of charged particle trajectories is achieved
primarily using the silicon tracker with both pixels and strips, embedded in
a 4 T magnetic field. The high granularity silicon pixel tracker consists of
three barrel layers (at about 4, 7 and 11 cm radius) and two endcap disks.
There are about 66 million pixels with an area of 100 $\times$ 150
$\mu\mathrm{m}^2$. The strip part is a combination of single- and
double-sided layers with ten barrel and nine forward layers on each side (9.3
million channels).  The reconstruction capabilities at lower $p_T$ are
limited by the high magnetic field and effects of the detector material. In
addition, in central Pb-Pb collisions the high occupancy of the silicon
strips makes the inclusion of these strips in charged particle tracking
difficult.

\begin{figure}[t]
 \begin{center}
 \includegraphics[width=0.49\textwidth]{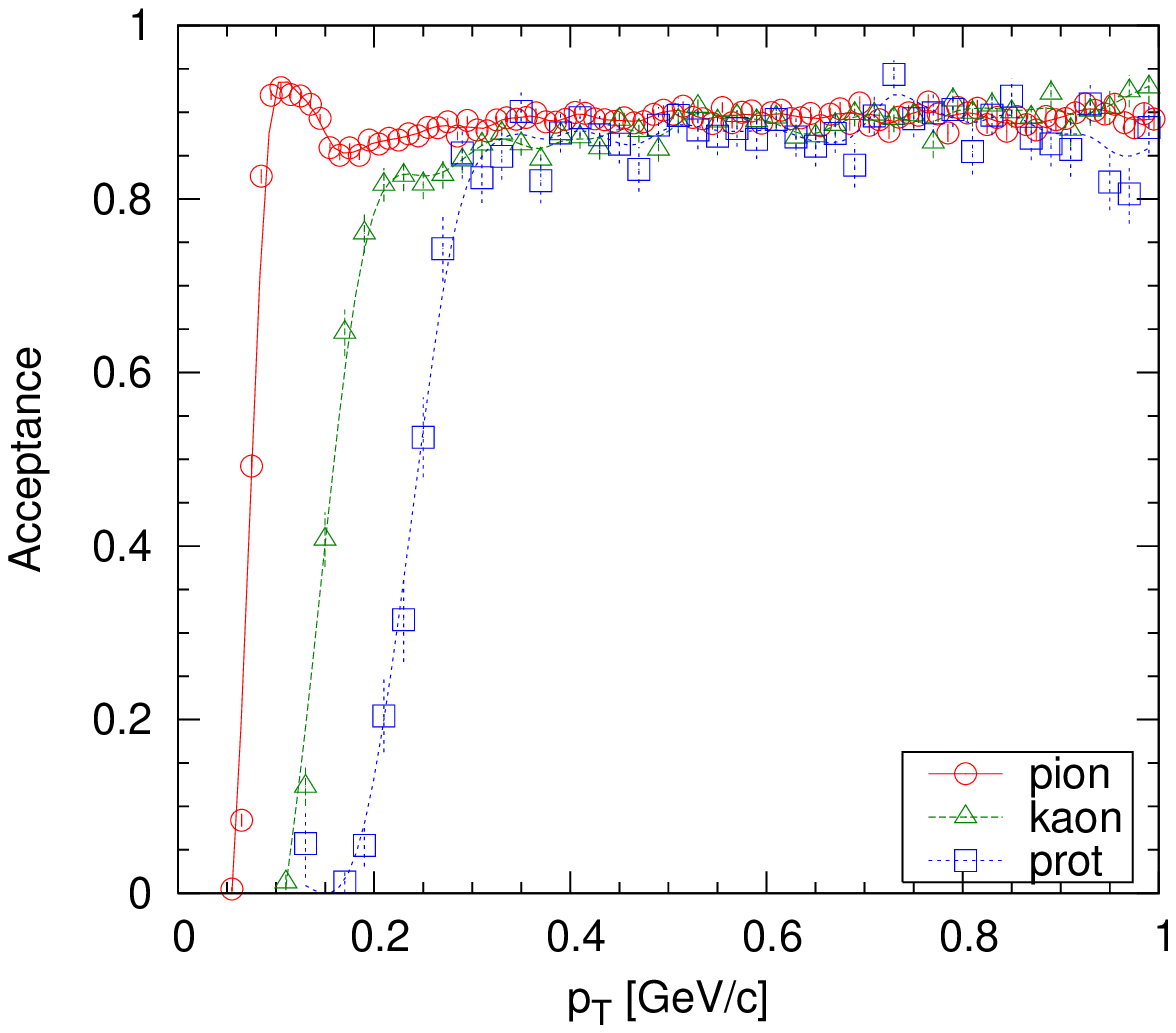}
 \includegraphics[width=0.49\textwidth]{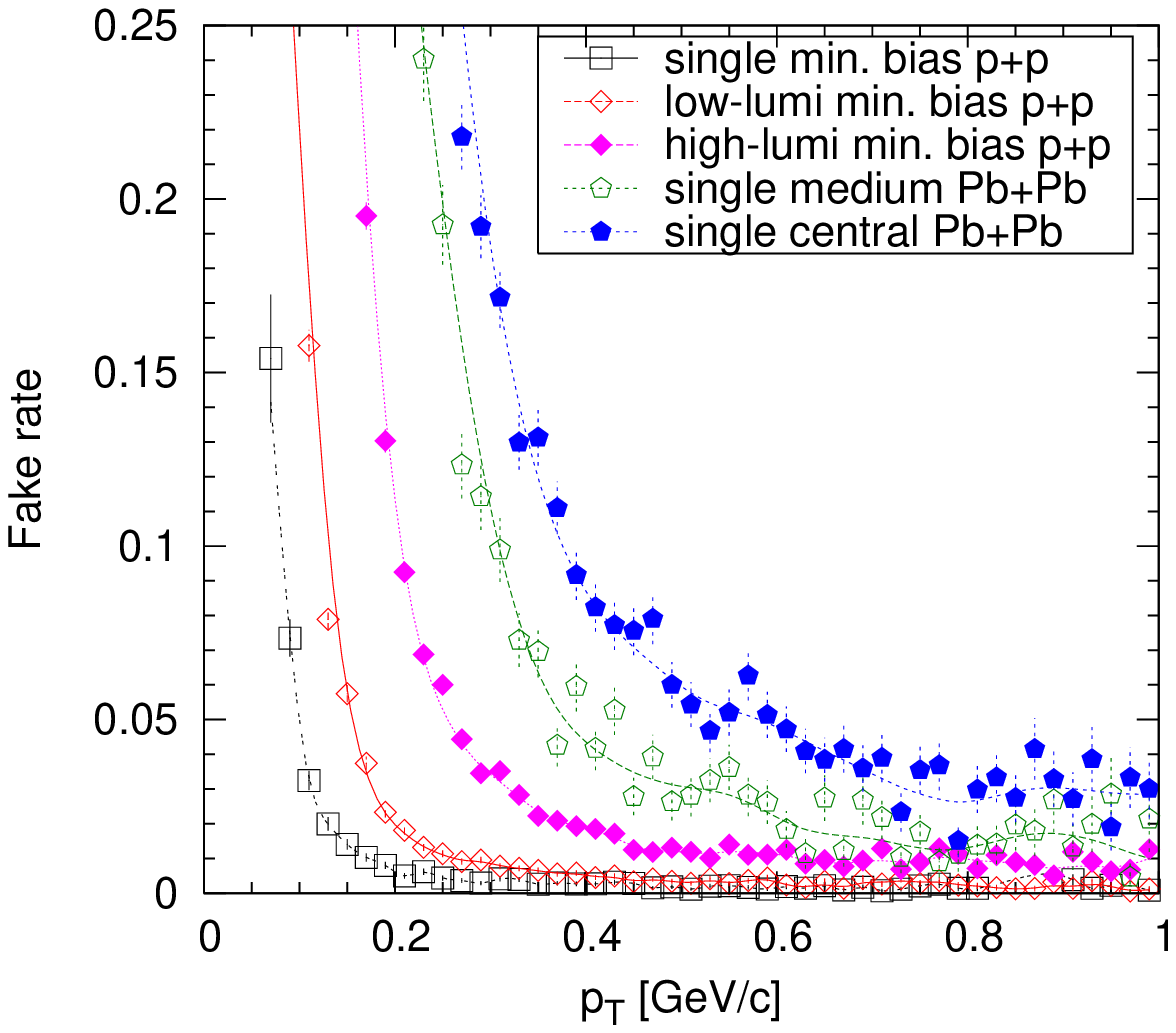}
 \end{center}

 \caption{Left: Acceptance of the track reconstruction algorithm as a
function of $p_T$ for tracks with $|\eta|<1$. Values are given
separately for pions (circles), kaons (triangles) and (anti)protons
(squares). Right: Reconstruction rate of fake tracks as a function of $p_T$,
for tracks with $|\eta|<1$, for single, low luminosity and high
luminosity minimum bias p-p events and for central and mid-central Pb-Pb
collisions. }

 \label{fig:lowpt}
\end{figure}

The track finding procedure, modified for low $p_T$, starts by pairing two
hits from different layers. During the search for the third hit, the following
requirements must be fulfilled: the track must come from the vicinity of the
beam-line; the $p_T$ of the track must be above the minimal value $p_{\rm
T,min}$; and the track must be able to reach the layer where the third hit may
be located.  In the small volume of the pixel detector the magnetic field is
practically constant and the charged particles propagate on helices. The
projection of a helix or a cylinder onto the transverse plane is a circle.
Each requirement defines a region of allowed track trajectories. They are
enclosed by a pair of limiting circles which can be constructed using simple
geometrical transformations. A third hit candidate is accepted if its position
is within a region which takes into account the expected multiple scattering.

While high $p_T$ tracks are relatively clean, uncorrelated hit clusters can
often be combined to form fake low $p_T$ tracks. A hit cluster contains more
information than just its position. Its geometrical shape depends on the
angle of incidence of the particle: large angles will result in longer
clusters. We can thus check whether the measured shape of the cluster is
compatible with the predicted angle of incidence of the track: if any of the
hits in the triplet is not compatible, the triplet is removed from the list
of track candidates.

Using this modified pixel hit triplet finding algorithm, charged particles
down to very low $p_T$ can be reconstructed (Fig.~\ref{fig:lowpt}-left).  The
acceptance and efficiency is at 80--90\%, while the $p_T$ resolution is about
6\%.  At the same time a low fake track rate is achieved thanks to the
geometrical shape of the hit cluster. It is below 10\% even in central Pb-Pb
collisions for $p_T>0.4$ GeV/$c$ (Fig.~\ref{fig:lowpt}-right). It is possible
to study identified particle spectra (down to $p_T$ of $0.1-0.3$ GeV/$c$)
and yields, multiplicity distributions and correlations \cite{hiptdr}.

\begin{figure}[t]
 \begin{center}
 \includegraphics[width=0.49\textwidth]{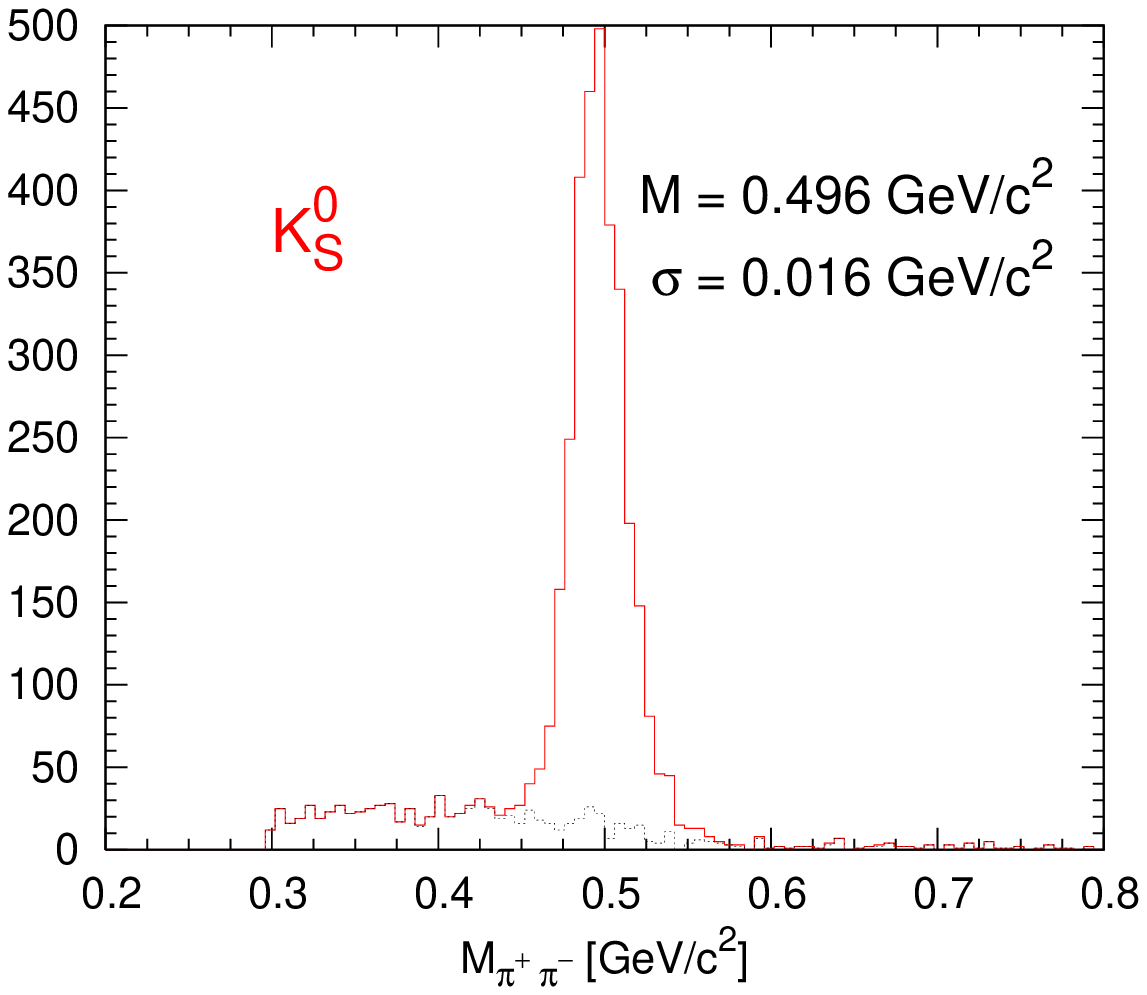}
 \includegraphics[width=0.49\textwidth]{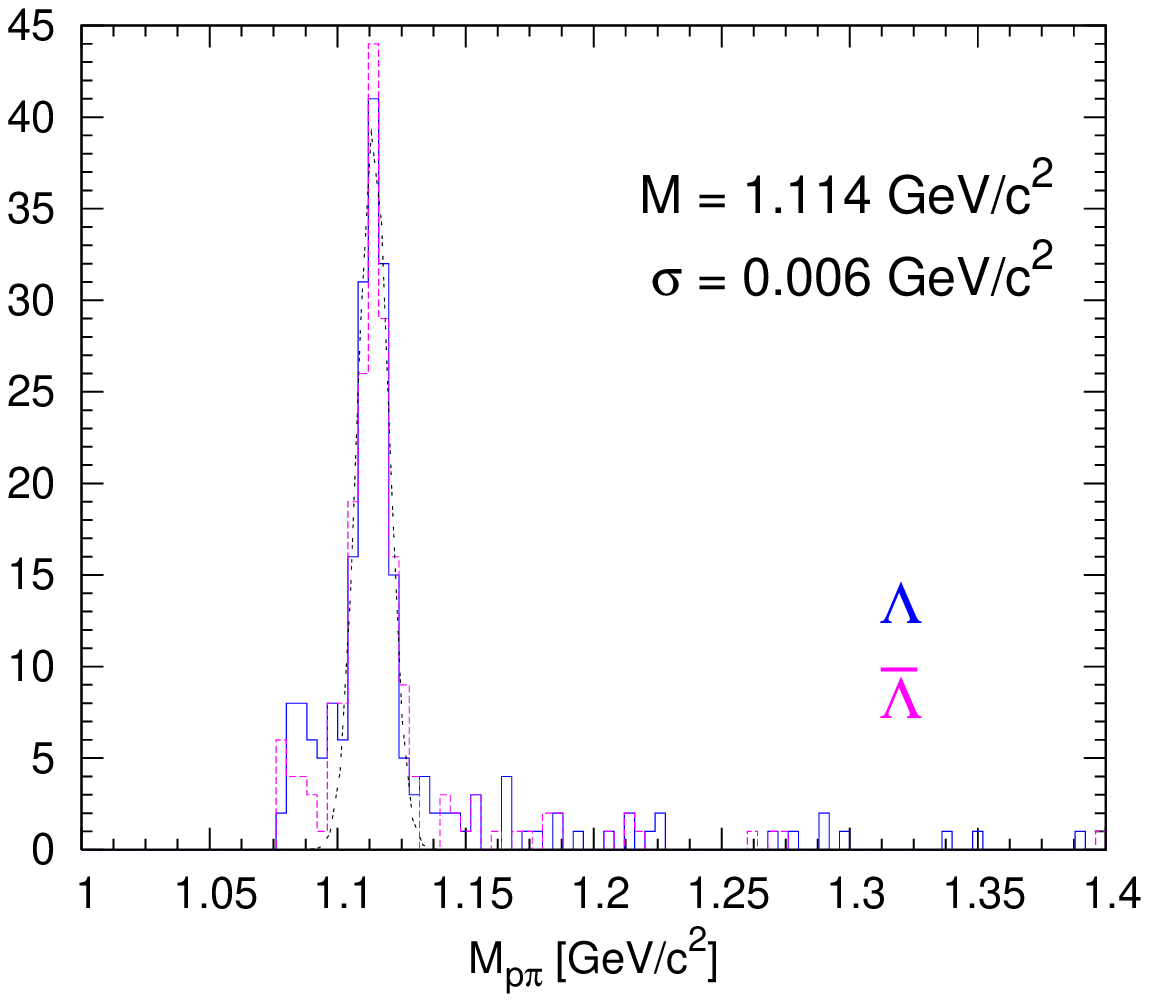} 
 \end{center}

 \caption{Left: Invariant mass distribution of reconstructed
$\mathrm{K^0_S}\rightarrow \pi^+\pi^-$ in single minimum bias p-p collisions.
The mass distribution of the background is indicated with a dashed
histogram. Right: Invariant mass distribution of reconstructed $\Lambda
\rightarrow \mathrm{p}+\pi^+$ and $\overline{\Lambda} \rightarrow
\mathrm{\overline{p}}+\pi^-$ in single minimum bias p-p collisions with
$dE/dx$ selection for the secondary proton or antiproton.}

 \label{fig:vzero}
\end{figure}

\subsection{Identified particles}

Particles can be identified by decay topology and by energy loss measurements
in the silicon (in the range $p<1-2$ GeV/$c$, benefitting from analogue
readout). Weakly-decaying neutral particles have a sizeable probability to
decay far from the primary event vertex. Likewise, the silicon detectors can
be used to reconstruct photons through their conversion to $\mathrm{e^+e^-}$
pairs in the material of the beam-pipe, silicon pixels and supports.  The
search for V0 candidates is reduced to determining the closest distance
between two track helices.  The $\mathrm{K^0_S}$ is reconstructed with a
resolution of 16~MeV/$c^2$ and an average mass of 0.496~GeV/$c^2$
(Fig.~\ref{fig:vzero}-left).  The $\Lambda$ and $\overline{\Lambda}$ peaks
are located at 1.114~GeV/$c^2$ with a 6~MeV/$c^2$ resolution
(Fig.~\ref{fig:vzero}-right).  The proton signal was strongly enhanced by a
cut on the truncated mean of their $dE/dx$, removing almost all the
background. In the case of single collisions or low luminosity p-p running,
the resonances can be exclusively identified. For high luminosity p-p running
or Pb-Pb collisions, the inclusive yield can still be extracted.
Multi-strange baryons ($\Xi^-$, $\Omega^-$), open charm ($\mathrm{D^0}$,
$\mathrm{D^{*+}}$) and open beauty ($\mathrm{B} \rightarrow \mathrm{J/\psi} +
\mathrm{K}$) can also be studied.

\subsection{Elliptic flow}

\begin{figure}[t]
 \begin{center}
 \includegraphics[width=0.49\textwidth]{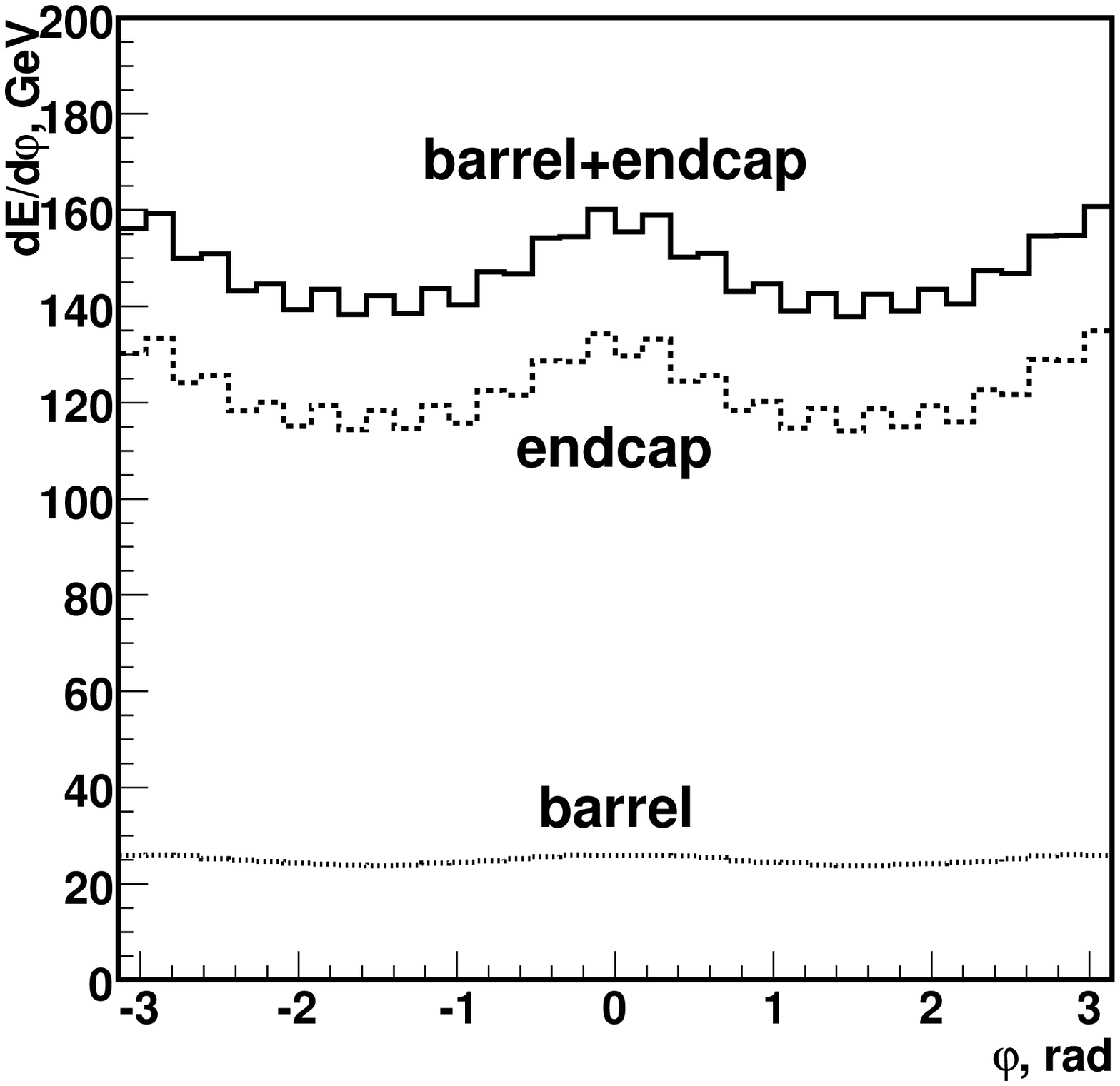}
 \includegraphics[width=0.48\textwidth]{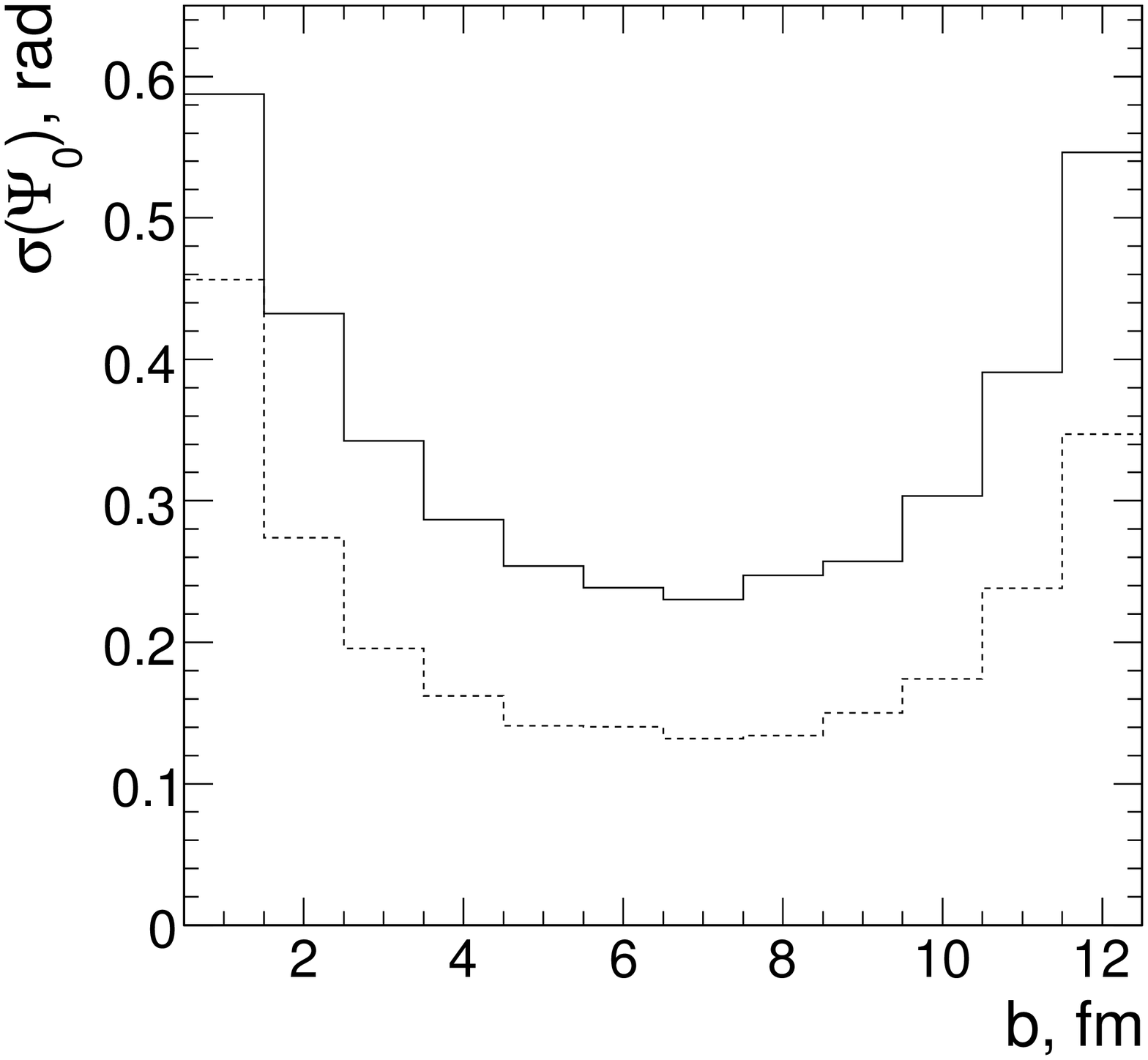}
 \end{center}

 \caption{Left: Azimuthal dependence of the total energy deposition in the
CMS calorimeters in Pb-Pb collisions at $b=9$~fm (solid histogram).  The
barrel and endcap regions are shown separately.  Right: Resolution of the event
plane angle $\Psi _0$, $\sigma (\Psi _0)$,  as a function of impact
parameter in Pb-Pb collisions with total particle multiplicities $N_0$~($b =
0$~fm)~=~58\,000 (solid histogram) and 84\,000 (dashed histogram).}

 \label{fig:energy-flow}
\end{figure}

The reaction plane in Pb-Pb collisions can be reconstructed using
electromagnetic and hadronic calorimeters, by extracting harmonic angular
coefficients of the energy deposition distribution $dE/d\varphi$
(Fig.~\ref{fig:energy-flow}-left).  The calculated event plane resolution as
a function of impact parameter is shown in Fig.~\ref{fig:energy-flow}-right.
The interplay of multiplicity and anisotropic flow in opposite centrality
directions gives the best resolution in semi-central
collisions.  The results will improve by adding tracker information and using
forward detectors such as the zero degree calorimeter. The 
second harmonic coefficient $v_2$
can also be determined using the two-particle
azimuthal correlator, without the event plane angle.  The
advantage of this method is that it automatically corrects for the detector
anisotropies.

\section{Summary}

The CMS detector has a good capability for global event characterization
and physics with soft probes. The performance in minimum bias triggering,
centrality determination, measurement of charged particle multiplicity, low
$p_T$ tracking, particle identification and measurement of elliptic flow has
been shown.

\section*{Acknowledgment}

The author wishes to thank to the Hungarian Scientific Research Fund (T
048898).

\end{document}